\documentclass[structabstract]{aa}
\usepackage{graphicx}
\usepackage{txfonts}
\usepackage{natbib}

\newcommand{\bt}{$B/T$}

\newcommand{\mstar}{$M_{\rm bulge}$}
\newcommand{\mbulge}{$M_{\rm bulge}$}
\newcommand{\msune}{M_{\odot}}
\newcommand{\mstaret}{M_{\rm star}}

\newcommand{\mbh}{$M_{\rm bh}$}
\newcommand{\mbhe}{M_{\rm bh}}

\newcommand{\sis}{${\sigma}$}
\def\msun{${\rm M_{\odot}}$}

\def\vmax{$V_{\rm max}$}

\begin{document}
   \title{Black Holes in Pseudobulges: demography and models}

   \subtitle{}

   \author{F. Shankar
          \inst{1}
          \and
          F. Marulli\inst{2,3,4}
          \and
          S. Mathur\inst{5,6}
          \and
          M. Bernardi\inst{7}
          \and
          F. Bournaud\inst{8}
          }

   \institute{GEPI, Observatoire de Paris, CNRS, Univ. Paris Diderot, 5 Place Jules Janssen, F-92195 Meudon, France\\
              \email{francesco.shankar@obspm.fr}
         \and
             Dipartimento di Astronomia, Universit\'{a} degli
Studi di Bologna, via Ranzani 1, I-40127 Bologna, Italy
\and
INAF-Osservatorio Astronomico di Bologna, Via Ranzani 1, 40127, Bologna, Italy
\and
INFN/National Institute for Nuclear Physics, Sezione di Bologna, viale Berti Pichat 6/2, I-40127 Bologna, Italy
   \and
   Department of Astronomy, Ohio State University, McPherson Laboratory,
140 W. 18th Ave., Columbus, OH 43210-1173, USA
\and
 Center for Cosmology and Astropharticle Physics, Ohio State
University, Columbus, OH 43210, USA
\and
 Department of Physics and Astronomy, University of Pennsylvania, 209
South 33rd St, Philadelphia, PA 19104
\and
 CEA, IRFU, SAp, 91191 Gif-sur-Yvette, France
             }

   \date{}


  \abstract
   {There is mounting evidence that a significant fraction of Black Holes (BHs) today live in late-type galaxies, including bulge-less galaxies and those hosting pseudobulges, and are significantly undermassive with respect to the scaling relations followed by their counterpart BHs in classical bulges of similar stellar (or even bulge) mass.}
   {Here we discuss the predictions of two
state-of-the-art hierarchical galaxy formation models in which BHs grow
via mergers and, in one, also via disk instability. Our aim is to understand if the wealth of new data on local BH demography is consistent with standard models.}
   {We follow the merger trees of representative subsamples of BHs and compute the fractional contributions of different processes to the final BH mass.}
   {We show that the model in which BHs always closely follow the growth of their host bulges, also during late disk instabilities (i.e., bars), produces too narrow a
distribution of BHs at fixed stellar mass to account for the numerous low-mass BHs now detected in later-type galaxies. Models with a looser connection between BH growth and bar instability instead
predict the existence of a larger number of undermassive BHs, in better agreement with the observations.}
   {The scatter in the updated local BH-bulge mass relation (with \emph{no} restriction on galaxy type) appears to be quite large when including later-type systems, but it can still be managed to be reproduced within current hierarchical models. However, the fuelling of BHs during the late bar-instability mode needs to be better quantified/improved to properly fit the data. We conclude discussing how the possibly large number of BHs in later type galaxies demands for an in-depth revision of the local BH mass function and its modelling.}


   \maketitle
%

\section{Introduction}
\label{sec|intro}

   There is evidence in the local Universe that most
bulge-dominated galaxies seem to obey rather tight correlations between
the central black holes' (BHs) masses and the bulge masses and velocity dispersions of
their hosts \citep[e.g.,][]{Magorrian98,Ferrarese00}.

Such tight correlations have been physically interpreted as byproducts of
some BH self-regulation mechanism -- active galactic nuclei (AGN) feedback -- operating
during the high-redshift, gas-rich phases of galaxy build-up.
At early times in fact, large amounts
of gas inflows can be generated due to, e.g., merger events or internal
instabilities \citep[e.g.,][]{Barnes92,Frederic11}, triggering star formation and
fuelling the central BH that can energetically and kinematically
back-react on the gas reservoirs via AGN winds.
Such interactions can possibly contribute to shut-down starformation
and to the formation of the progenitors of today's early-type galaxies
\citep[e.g.,][]{Sanders88,Granato06,Shankar08,Lapi11}.

Mounting empirical evidence suggests, however,
that other additional tracks may have significantly contributed to BH
formation and growth.
Different observations now question that BH-host
correlations indeed appear as tight as previously claimed for all types of galaxies
\citep[e.g.,][]{Gultekin09,Graham09barred}. Significant outliers exist
\citep[e.g.,][]{Mathur01,Grupe04,Watson07}, BHs are currently observed also in
the centres of bulge-less galaxies \citep[e.g.,][]{Ghosh08,Satyapal09,Araya12},
and the AGN--starformation correlations disappear for all but
highest luminosity AGNs \citep{Shao10,Grier11}.

Perhaps one of the most pressing reason to argue for supplementary routes of BH evolution is the prevalence of pseudobulges hosting BHs
\citep[e.g.,][]{Hu08,GadottiKauffmann,Kormendy11,Mathur11,Orban11}.
``Pseudobulges'' are structurally and dynamically different from
``classical'' bulges and ellipticals, which can also have very low specific angular momentum \citep[e.g.,][]{Emsellem11}. They appear to have more ``disky''
profiles, they depart from the fundamental plane of elliptical galaxies,
and are fainter at a fixed effective radius \citep[e.g.,][]{Gadotti09}.

   \begin{figure*}
   \centering
   \includegraphics[width=18truecm]{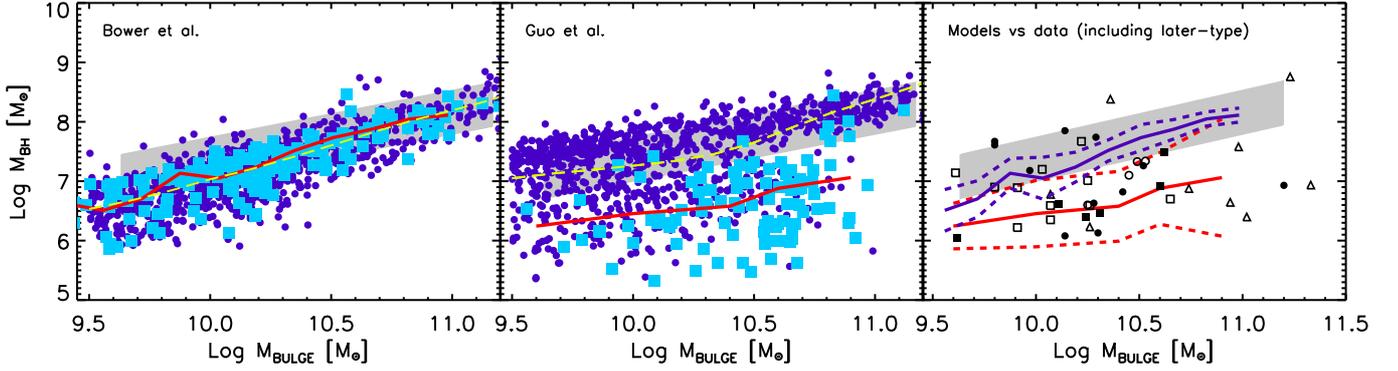}
   \caption{\emph{Left panel}: The \emph{dark blue circles} show the
    distribution of black holes for a sample of 1000 black holes
    randomly selected from the \citet{Bower06} catalogue. The \emph{cyan
    squares} are a random sample of 100 black holes with their hosts'
    bulges mainly grown via secular processes ("pseudobulges"). The
    \emph{long-dashed yellow} and \emph{red black} lines are the median
    \mbh-\mstar\ relations competing to the former and the latter
    samples, respectively; the \emph{grey stripe} indicates the local
    empirical \mbh-\mbulge\ relation for classical bulges from \citet{Sani11}.
    \emph{Middle panel}: same pattern
    as the left panel but with both samples extracted from the
    \citet{Guo11} catalog, with the \emph{long-dashed yellow} and
    \emph{red solid} lines being the respective median \mbh-\mstar\
    relations for the two samples.  \emph{Right panel}: comparison
    between predicted the distributions of pseudobulges of
    \citet{Bower06} and Guo et al. (2011; \emph{blue} and \emph{red} lines,
    respectively), with their 1-$\sigma$ uncertainties (\emph{dashed}
    lines) compared to data (Table 1) from \citet[][\emph{filled
    squares}]{Hu09}, \citet[][\emph{filled circles}]{Kormendy11},
    \citet[][\emph{open circles}]{Mathur11}, and \citet[][\emph{open
    squares}]{Orban11}. Also shown for completeness data on pseudobulges
    by \citet[][\emph{open triangles}]{Sani11}. The average error in
    BH masses for active and inactive galaxies is
    of the order of $\sim 0.3$ dex. Of similar magnitude is the propagated error on stellar masses
    at fixed galaxy colour.}
    \label{fig|MbhMstarRelation}
    \end{figure*}

At variance with elliptical galaxies and classical bulges, pseudobulges
have been suggested to have formed through secular processes such as
disk instabilities (\citealt{Kormendy04}).
What we refer to here is \emph{bar} instability, physically distinct to the clumpy instabilities that characterize the
high-$z$, gas-rich disks \citep[e.g.,][]{Daddi10,Genzel11}.

Indeed, \citet{Shankar11}
have recently shown that bulges mainly formed from bar-like disk instability in the \citet{Guo11} model have much lower bulge-to-total (B/T) ratios and smaller sizes than classical bulges at fixed total stellar
mass. \citet{Hopkins10bulge} have argued that the formation of
pseudobulges depends on how gas-rich the merger is, and
\citet{Fontanot11} claim that it depends on the mass of the dark matter
halo (see also discussions in \citealt{Fanidakis11} and \citealt{Yates11}).
In particular, as detailed below, BHs residing in pseudobulges
appear to be relatively undermassive compared to those in classical
bulges of similar luminosity or velocity dispersion
\citep[e.g.,][]{Hu08}.


Despite the strong observational efforts carried out over the years
to characterize the BH population in later-type galaxies, theoretical models have mostly
focused their attention on BHs in classical bulges.
Here we investigate the formation of late-type bulges and their
nuclear BHs, with emphasis on the class of galaxies broadly referred to as pseudobulges, in the framework of state-of-the-art hierarchical galaxy
formation models that grow bulges and/or their central BHs via mergers and disk
instabilities. In particular, we will show that mergers can produce a
significant scatter in the BH-galaxy mass scaling relations,
in broad agreement with the observed properties of classical and pseudo bulges.

\begin{figure*}
    \centering
    \includegraphics[width=17truecm]{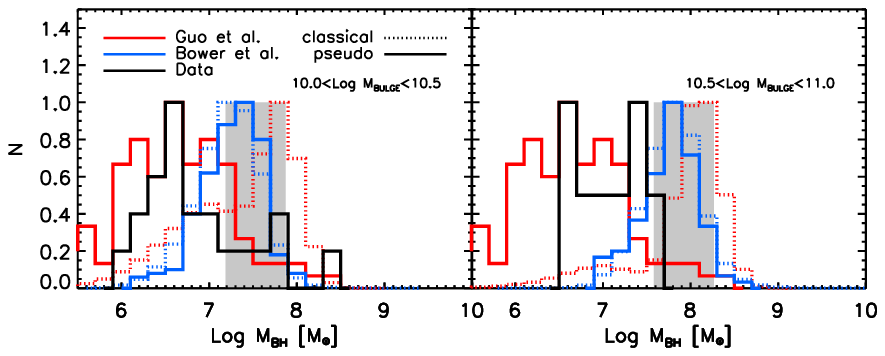}
    \caption{Black hole mass distributions for galaxies with bulge masses $10<\log M_{\rm bulge}<10.5$ (\emph{left panel}) and
    $10.5<\log M_{\rm bulge}<11$ (\emph{right panel}) predicted by the \citet{Guo11} and \citet{Bower06} models (\emph{red} and \emph{blue} lines, respectively), and for classical and pseudo bulges (\emph{dotted} and \emph{solid} lines, respectively). The solid, black lines in both panels are the data collected in Table~\ref{table|models}. The \emph{grey stripes} are the black hole mass intervals (median plus scatter) inferred from the Sani et al. \mbh-\mstar\ relation for the corresponding bulge masses.}
    \label{fig|Scatter}
\end{figure*}

\section{Reference Models}
\label{sec|Model}

The results discussed here are based on the online catalogues of
the \citet[][B06 hereafter]{Bower06} and the \citet[][G11 hereafter]{Guo11}
semi-analytic models \footnote{http://gavo.mpa-garching.mpg.de/MyMillennium3}.

Both models follow the evolution of galaxies and their central BHs along the hierarchical evolution
of their host dark matter haloes
and subhaloes within the concordance $\Lambda$CDM cosmology. To this purpose, they are both
implemented on top of the \emph{same} large, high-resolution cosmological N-body simulation
{\tt MILLENNIUM I} \citep{Springel05}, making their comparison even more
meaningful. The {\tt MILLENNIUM I} simulation follows the evolution of $N= 2160^3$ dark matter particles of mass
$8.6\times10^{8}\,h^{-1}{\rm M}_{\odot}$, within a comoving box of
size $500\, h^{-1}$Mpc on a side, from $z=127$ to the present, with cosmological parameters
$\Omega_{\rm m}=0.25$, $\Omega_{\rm b}=0.045$, $h=0.73$,
$\Omega_\Lambda=0.75$, $n=1$, and $\sigma_8=0.9$. The simulation can
resolve haloes down to a mass of $\gtrsim 10^{10}\, M_{\odot}/h$, corresponding
to BHs close to the seed mass used in the models ($\gtrsim 10^4\, M_{\odot}$),
thus, any conclusion drawn for BHs more massive than $\sim 10^5\, M_{\odot}$
is free of any resolution effect.

The primary channel to form bulges in both models is via mergers,
during which bulges gain a fraction of the progenitor stars along with
newly formed stars, both dependent on the strength and type of the merger.

Both models also substantially grow bulges via bar-like disk instabilities.
Secular evolution is believed to form bulges via a bar that is capable of transferring mass from the disk to the central regions. At each timestep for each galactic system a stability criterion of the type
\begin{equation}
V_{\rm max} < V_{\rm ref}\, ,
\label{eq|vmax}
\end{equation}
is checked.  Here \vmax\ is the velocity at the maximum of the rotation
curve, while $V_{\rm ref} \propto \sqrt{GM_{\rm disk}/R_{\rm disk}}$ is
proportional to the actual velocity of the disk via a factor close to
unity.

When Eq.~\ref{eq|vmax} is satisfied the models assume that the
instability drives the formation of a bar from the disk into a central
bulge \citep[e.g.,][]{Combes90,Cole00}.  Despite the broad scheme of the
disk instability is apparently similar in both models, there are
significant differences. G11 treat disk instabilities as secular
processes that act in self-gravitating disks near the instability
boundary by transferring a portion of the stellar mass $\delta \mstaret$
from the disk to the bulge to keep the disk marginally stable with an
exponential density profile.

B06 instead assume that the entire mass of the disk is transferred to
the galaxy bulge when the disk goes unstable, with any gas present
assumed to undergo a starburst and a fraction of it to feed the BH. It is thus expected
that disc instabilities play a more prominent role in the B06 model. B06
indeed find that in their model most of the BH growth occurs via disk
instabilities rather than mergers.  The disk instabilities considered by
B06 may also qualitatively share some features with the instabilities
mentioned in Sect.~\ref{sec|intro} characterizing high-$z$, gas-rich
disks \citep[e.g.,][]{Genzel11}, in the sense that in the B06 a
significant fraction of gas can be fuelled to the centre even during
disk instabilities. Nevertheless, the B06 are still considered and
modelled as bar-like instabilities accompanied by a starburst (see
discussion in \citealt{Cole00}), thus still quantitatively different
from the clumpy ones discussed in Sect.~\ref{sec|intro}. Galaxy
evolution models still lack a proper treatment of bulge formation (and
any consequent BH feeding) via clump migration in gas rich, turbulent
systems (see further discussions in Sect.~\ref{sec|discu} and in,
e.g., \citealt{Dekel09}).

\begin{figure*}
    \centering
    \includegraphics[width=18truecm]{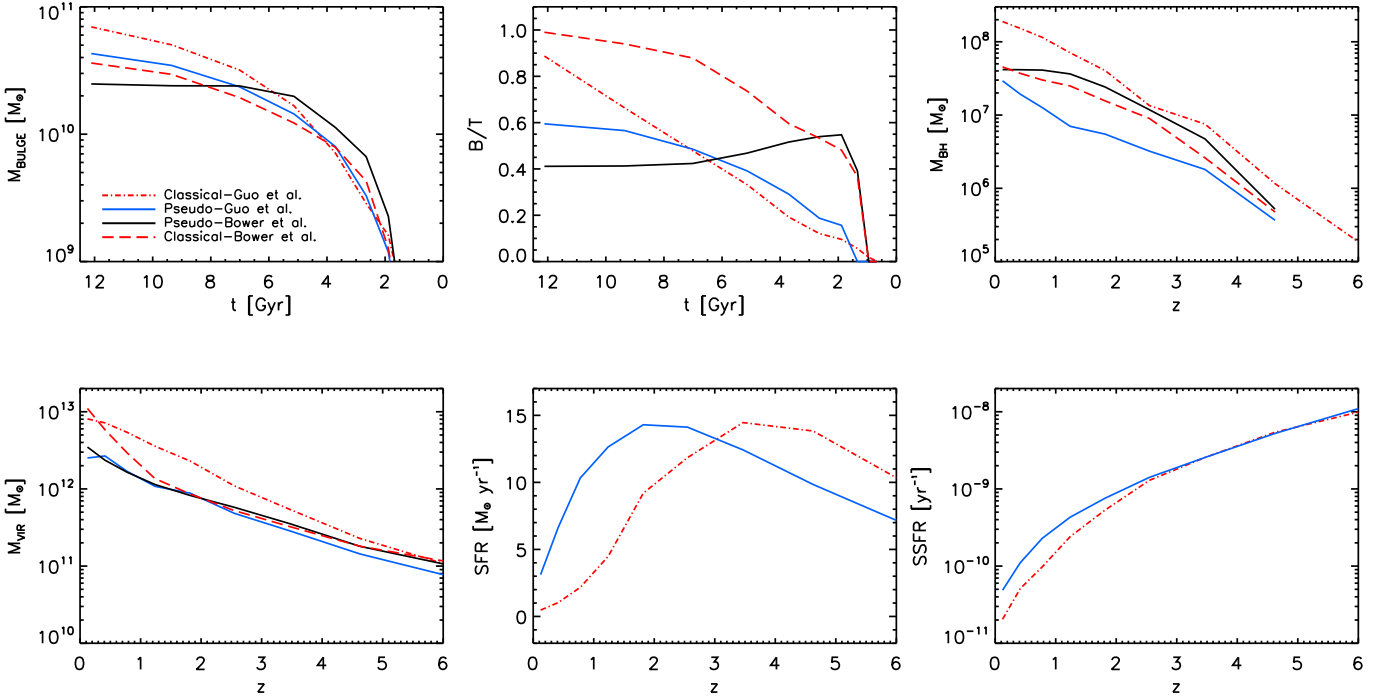}
    \caption{Comparison among the median properties of random subsamples
    of classical and pseudobulges from the \citet{Guo11} model
    (\emph{red}, \emph{dot-dashed} and \emph{blue}, \emph{solid} lines,
    respectively) and \citet{Bower06} model ( \emph{red},
    \emph{long-dashed} and \emph{black}, \emph{solid} lines,
    respectively). \emph{Upper panels}: bulge mass (\emph{left}),
    bulge-to-total \bt\ ratio (\emph{middle}), and black hole mass
    (\emph{right}) as a function of cosmic time or redshift.
    \emph{Lower panels}: subhalo virial host mass (\emph{left}),
    star-formation rate (\emph{middle}), and specific star-formation
    rate (\emph{righ}t) as a function of redshift. Overall, pseudobulges
    appear to have prolonged star-formation rates, significantly lower
    \bt, and in models with only mergers as fuelling mechanism, lower
    mass black holes, in agreement with observational data.}
    \label{fig|Hosts}
\end{figure*}

Both models postulate that at each merger the central BH accretes mass
proportionally to the available gas in the host galaxy (in the G11 model
the accretion efficiency gets lower for smaller mass systems and for
unequal mergers), as well as mass from other incoming BHs.

The most noticeable difference relevant to the present work is that, at variance
with the G11 model, BHs in the B06 model also grow in mass during disk
instabilities, proportionally to the fraction of gas available and
undergoing a starburst at the moment of the disk instability.
That some of the gas can be channelled into the very central
regions during disk instabilities echoes some of the earlier models
such as the so-called ``bars within bars'' scheme,
proposed as a possible mechanism for fueling AGNs \citep[e.g.,][]{Shlosman89}.
For suitable conditions the gas inflow induced by the bar instability
may produce a self-gravitating disk in a smaller region and
a new small bar can develop, which induces further gas inflow.

In both models BHs also slightly grow in mass via hot
accretion at late times (radio-mode feedback), but this is a minor
channel of BH growth compared to cold gas accretion and BH-BH mergers.
Anyway, including or neglecting such processes in the models is irrelevant
to the present discussion.

Both models have been fine-tuned in their input parameters to
reproduce the local scaling relations between BH mass and host galaxy stellar mass. However, the intrinsic \emph{scatter} in the local relations is a genuine prediction of the models. The scatter reflects the relative diversity in the growth histories of central BHs hosted by galaxies of similar mass, induced by the
intrinsic dispersion in the merger histories,
and the different underlying physical assumptions (mergers and/or disk instability) to grow BHs in the models.
Our discussion below will start from
the comparison of the predicted scatter against the wealth of data
now available on the BH-galaxy scaling relations.


\section{Data}
\label{sec|data}

In the following, we will compare model predictions with a collection of
data taken from the Literature, listed in Table 1. These
data are for active and inactive BHs the hosts of which have been
mostly recognized as pseudobulges. The column 3 of Table 1 lists references for
both BH and bulge masses. Where stellar masses were not directly
available, we obtained them by converting $K$, $r$, or $V$ magnitudes
using the mass-colour relations from \citet{Bell03} with an average
$g-r=0.5$, typical of these galaxies. All the data and the models have
been converted to a common \citet{Ch03} Initial Mass Function.  In
\citet{Kormendy11} BH masses have been measured using dynamical
methods. In \citet{Mathur11} and \citet{Orban11} widths
of the H$\beta$ emission lines in AGN spectra have been used for BH mass
estimates (see Mathur et al. 2011 for details. Orban de Xivry et al. do
not list the BH masses in their paper; these have been kindly supplied
by the authors). Hu (2009) and Sani et al. (2011) report BH masses from
multiple sources (please see their papers for details).

Different authors have used different techniques to estimate BH
(and bulge) masses, and each technique has its own set of errors. The average error in
BH masses for active and inactive galaxies is
of the order of $\sim 0.3$ dex (see, e.g., \citealt{Mathur11}).
Of similar magnitude is the propagated error on stellar masses
at fixed galaxy colour (see, e.g., \citealt{Bell03}).
In the following, for each galaxy we will use and show the multiple available
BH and relative bulge mass estimates evaluated by each group.
As it will be shown and discussed below,
while some of the scatter in the data is undoubtedly induced by different measurement
techniques, despite the different selections and
methodologies, these data share a similar portion of the \mbh-\mstar\ plane,
mainly localized \emph{below} the classical \mbh-\mstar\ relation. The latter
is not a new result and is discussed in the papers listed above. Here we
simply put these data together for comparison with models.

\citet{Greene08} and \citet{Jiang11} have confirmed the results of the above mentioned works, extending them to lower BH masses. \citet{Greene10} also report lower
BH masses at fixed velocity dispersion for a (smaller) sample of local
late-type galaxies with more accurate BH masses from maser kinematics.
However, we do not report their results in Table 1 as only one galaxy in their sample
with secure bulge masses is recognized as a pseudobulge.
Also the indirect arguments put forward by \citet{GadottiKauffmann} suggest
pseudobulges have BHs less massive than their classical counterparts at fixed
bulge mass or velocity dispersion.

\section{Results}
\label{sec|results}

\subsection{A narrow and a broad distribution of Black Holes}
\label{sec|BHsDistributions}

From now on we will simply refer to bulges grown mainly via bar-like disk instabilities as ``pseudobulges'', as opposed to ``classical'' bulges that have grown their mass mainly via mergers.
Such a classification is rather arbitrary, though refined
theoretical and observational work is now
establishing that movement of gas and stars via bars to the centre
could effectively contribute in building a young and kinematically cold stellar
bulge component, that is, a disc-like/pseudo bulge \citep[e.g.,][and references therein]{Gadotti11}.
Moreover, it was recently pointed out by \citet{Shankar11} that galaxies of
similar stellar mass that underwent a different merger history appear
structurally different in the local Universe. In particular, they
emphasized that in the G11 model those galaxies that had their bulges mostly grown via
disk instability end up having
significantly lower bulge-to-total ratios \bt\ and lower half-mass
radii at fixed stellar mass, in agreement with data \citep[e.g.,][]{Gadotti09}.
Here we extend the differences
emphasized by \citet{Shankar11} to BH masses and other properties.

The blue circles in the left panel of Fig. 1 are a random subsample of
1000 BHs extracted from the B06 catalog, with the yellow, dashed line marking the median relation.
A relatively narrow \mbh-\mstar\ relation is predicted in fair agreement
with what empirically calibrated in the local Universe (the grey band is
the \mbh-\mstar\ fit for classical bulges by Sani et al. 2011 with its
intrinsic scatter; similar results were found by, e.g.,
\citealt{HaringRix}, though the latter are mainly based on dynamical masses).
The cyan solid squares represent 100 randomly
extracted BHs hosted by pseudobulges, with the red, long-dashed line marking the median
relation. These galaxies were initially selected
from the catalogs as ``potential'' pseudobulges for having lower \bt\ (usually set to be \bt$\lesssim 0.5$)
at fixed \emph{bulge} stellar mass. By analyzing their merger trees, following
\citet{Shankar11} we then
selected and defined as pseudo (classical) bulges those galaxies that had grown more (less) than 50\% of their
bulge mass via disk-instability.
We repeated the exercise of randomly extracting 100 galaxies with such characteristics several times
finding similar results.  Given that BHs even in disk-instability mode
in the B06 model grow proportionally to bulges, it is not unexpected
that they will end up following similar scaling relations to the BHs in
classical bulges.


The middle panel of Fig. 1 shows the predictions of the G11 model. The
predicted \mbh-\mstar\ relation (blue circles), with the median shown with a yellow, long-dashed line,
is still broadly consistent with local data, though the predicted scatter at fixed
stellar mass is much larger in this model. The cyan squares are a
subsample of 100 randomly selected BHs that have grown more than 50\% of
their bulge mass via bar-like disk instability. Galaxies of the latter
type were again initially selected as having lower \bt\ at fixed bulge
mass. It is clear that extracting merger histories for the millions of BHs in the online catalogues becomes
prohibitive. Nevertheless, for the G11 model we compared our randomly selected pseudobulge sample
with one directly extracted from the \citet{Shankar11} catalog. The latter is a rendition of the G11
model with additional information on bulge sizes, where the ratio between size growth
via merger and bar instability was self-consistently followed for each galaxy in the model
(see further details in \citealt{Shankar11}). From such a catalog we randomly extracted 100
galaxies with bulge sizes mostly grown via bar instabilities finding very similar results compared to our
random samples.


In the right panel we compare model predictions with a collection of
data taken from the Literature, listed in Table 1 and discussed in Sect.\ref{sec|data}.
We stress that these
data are for active \emph{and} inactive BHs the hosts of which have been
recognized as pseudobulges. These data show that despite the different
selections and methodologies, they share a similar portion of the
\mbh-\mstar\ plane. The most striking feature
shown by this collection of data is that at fixed \emph{bulge} mass,
most of the objects are mainly localized \emph{below} the classical
\mbh-\mstar\ relation, characterized by the grey band, by up to a factor of
$\lesssim 100$ and over nearly two orders of magnitude in bulge stellar mass.
The blue and red lines
are the predictions for pseudobulges of the B06 and G11 models,
respectively, with the solid and dashed lines marking their respective
mean and 1-$\sigma$ uncertainties (computed from
the 16th and 84th percentile of the $\log \mbhe$ distributions
competing to each bin of bulge stellar mass).


We find that the B06 model predicts a rather tight \mbh-\mstar\ relation
for all galaxies with normalization consistent with that of classical
bulges. Thus the model can only marginally account for BHs in
later type galaxies that observationally mostly lie well below
the classical \mbh-\mstar\ relation. The G11 model performs better in this respect.  Although a
significant fraction of the local BH population clusters around the
classical \mbh-\mstar\ relation, a non-negligible portion extends to
very low mass BHs at fixed \mstar.  The solid and dashed blue lines
highlight the distribution of galaxies in the G11 model that have
grown most of their bulge mass via disk instabilities.  More precisely,
we find that moving, at fixed \mstar, from the lowest mass BHs to the
highest ones, the fraction of mass assembled via mergers gradually
increases. The merger model therefore does not predict a bimodality of
BHs in classical and pseudo bulges, but rather a \emph{continuous} trend.

A more quantitative comparison between models and data is provided in
Figure~\ref{fig|Scatter} which shows BH mass distributions for a fixed bin of bulge stellar mass.
Given that we are here mainly interested in the comparison
among predicted and measured scatters in BH mass at fixed bulge mass,
we simply normalize each histogram to have the same peak at 1.
Distributions for galaxies with bulge masses $10<\log M_{\rm bulge}<10.5$ and
$10.5<\log M_{\rm bulge}<11$ are plotted in the left and right panels, respectively.
The red and blue lines are the predictions
for the G11 and B06 models, respectively, separated for classical
and pseudo bulges (dotted and solid lines, respectively).
The solid, black lines in both panels are the data collected in Table~\ref{table|models},
while the \emph{grey stripes} are the black hole mass intervals (median plus scatter)
inferred from the \citet{Sani11} \mbh-\mstar\ relation for the corresponding mean bulge masses.

For each $\log \mbhe$ distribution we then compute the median value,
and the 16th and 84th percentile which define the $\pm 1\sigma$ errors on $\log
\mbhe$. We find that, at least above $\log M_{\rm bulge} \gtrsim 11$
both models predict a narrower distributions for classical bulges, with an intrinsic
$\sigma(\log \mbhe) \sim 0.3-0.4$ dex, in good agreement with the empirical one
of 0.38 dex (grey band). However, while the B06 model continues to predict a similarly narrow
distribution for pseudobulges, the G11 model produces much broader distributions
for pseudobulges, with an intrinsic scatter of $\sigma(\log \mbhe) \sim 0.6$ dex,
a factor of about two higher than for the B06 model and in better agreement with the data
(cfr. solid, red lines, the G11 model/pseudo, with the black, solid histograms, the data).
At lower bulge masses (left panel), both models envision similar BH mass
distributions and similar differences, with the G11 model better matching observations.

In the G11 model seed BHs (set to $10^3 \msune$) grow via mergers by
orders of magnitude to the level of $\gtrsim 10^5-10^7 \msune$,
comparable to the BH masses measured in late-type galaxies. At later times,
when mergers become rarer, BH growth is blocked while host galaxies
continue to grow via in-situ star formation and bulges grow via disk
instabilities.  Thus the predicted \mbh-\mstar\ relation in the G11
model will inevitably show a rather large scatter at fixed bulge mass.


\subsection{Pseudobulges and Classical bulges: Evolution}
\label{sec|DifferentHosts}

In Fig. 3 we show several mean properties for the subsamples of
pseudobulges of the B06 and G11 models discussed earlier (solid black
and blues lines, respectively), compared with the mean properties of
subsamples of classical bulges (red long-dashed and dot-dashed
lines, respectively). For each object we trace back its more massive
progenitor in the merger tree and record its properties.
\citet{Shankar11} found that the sizes of pseudobulges are predicted to
be a factor of a few lower than those of classical bulges at fixed
stellar mass. Fig. 2 extends their study to other properties.

\subsubsection{$B/T$ versus $z$}

We find that the bulge-to-total \bt\ ratios (upper middle) of both galaxy
populations increase rapidly at very early epochs. However, while
classical bulges continue to significantly grow and turn into
ellipticals (\bt$\sim 1$) at late times, the \bt\ of pseudobulges
quickly flatten out to values \bt$\lesssim 0.6$ after a few Gyrs, as a
consequence of their significantly different growth patterns
\footnote{The apparent slight decrease in the \bt\ for pseudobulges in the B06 model
is simply an artifact of the averaging, the large intrinsic scatter, and the choice
of galaxy subset, but not driven by any evident physical process.}.

\subsubsection{$SFR$ versus $z$}

Pseudobulges in both models reside in lower mass haloes at $z\lesssim 2$
(lower left), which contributes in explaining the different merger histories; it also
induces less powerful mergers, thus less BH growth at all times.
We find in fact that pseudobulges have grown $>70\%$ of their final stellar
mass via disk instabilities (see also discussion in B06), while classical bulges of the same mass
tend to grow mostly via late dry minor mergers \citep[see details
in][and references therein]{Shankar11}.  This is also evident from their
different cosmological star formation rate (SFR; lower middle;
information on the SFR are only available for the G11 catalogue) and
specific SFR (SSFR; lower right).  The SFR in pseudobulges peaks at
later times and their SSFR is larger by a factor of a few at $z \lesssim
2.5$ with respect to classical bulges that become progressively dead
with time though still significantly increasing in stellar mass (via dry
mergers).

\subsubsection{BH mass versus $z$}

The characteristic BH mass (upper right) in pseudobulges for
the G11 model (solid blue) is a factor of several lower
($\gtrsim 4--10$ depending on redshift) than those in
classical bulges (red dot-dashed) since early times, at variance with
the bulge masses that differ by only a factor $\lesssim 2$ (upper
left). The B06 model instead predicts that both BH and bulge masses of
pseudo (solid black) and classical (red long dashed) bulges are similar
because they always co-evolve.  At late times pseudobulges continue
forming stars in their disks, undergoing disk instabilities, and growing
their bulges. However BHs in the G11 model suffer severe starving due to
less numerous and less powerful merger events, thus inducing further
departure from the classical \mbh-\mstar\ relation\footnote{We note that each subset of mock galaxies was mostly chosen on the basis of being enough massive to host a super-massive BH, and on the relative fractional growth via mergers and disc instabilities. We thus expect by construction the pseudo and classical bulge subsets within each model to yield similar stellar masses at $z=0$ (at the factor of $\lesssim 2$ level). However, we do not expect the classical and/or pseudobulge subsamples between the two model to necessarily yield the same bulge, BH, or virial mass at $z=0$.}. We'd like to conclude stressing that what discussed here are possibly \emph{extreme}
BHs in pseudo and classical bulges, but the models also predict
significant fractions of BHs with intermediate properties.

\subsection{Late-type bulges: BH demography revisited}
\label{sec|BHMFpseudo}

\begin{figure*}
    \centering
    \includegraphics[width=17truecm]{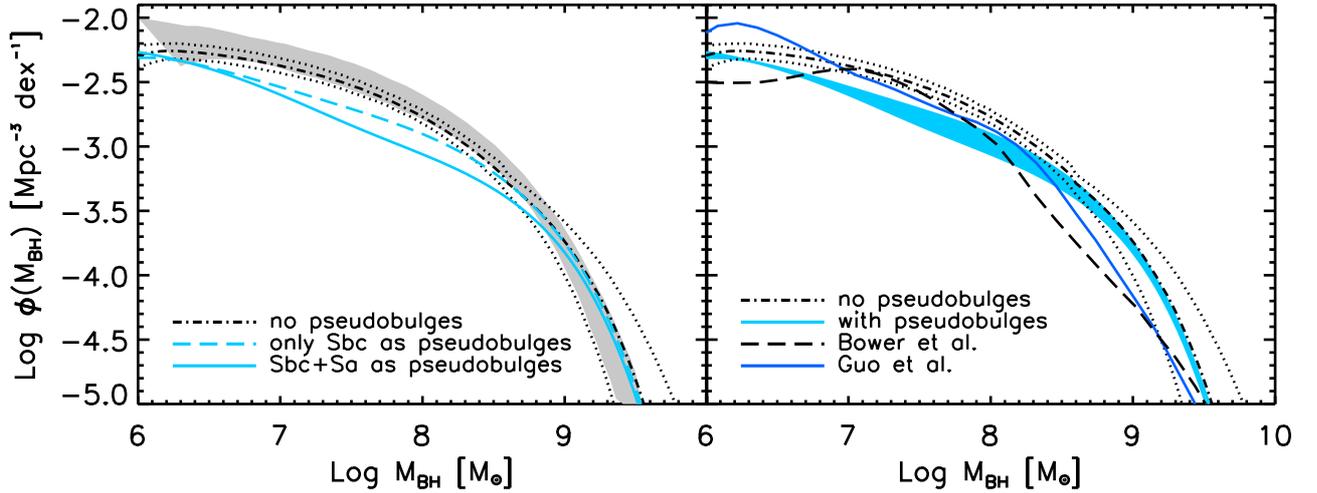}
    \caption{\emph{Left panel}: estimate of the local black hole mass
    function derived by convolution of the galaxy velocity dispersion
    function and the \mbh-\sis\ relation considering all bulges as
    classical (\emph{dot-dashed} and \emph{dotted} lines), compared to
    the collection of estimates from \citet[][\emph{grey
    band}]{SWM}. The \emph{cyan solid} line is the estimate of the local
    black hole mass function assuming that all Sbc+Sa galaxies follow
    the \citet{Hu09} \mbh-\sis\ relation proper of pseudobulges. The
    \emph{cyan dot-dashed} line is the local black hole mass function
    obtained assuming that only Sbc are pseudobulges. \emph{Right
    panel}: predicted local black hole mass function from the
    \citet{Bower06} and \citet{Guo11} models (\emph{black long-dashed}
    and \emph{blue solid} lines, respectively) compared with the local
    mass function not corrected (\emph{dot-dashed} and \emph{dotted}
    lines) and corrected for pseudobulges (\emph{cyan area}, combination
    of the \emph{dot-dashed} and \emph{solid} lines in the left panel).}
    \label{fig|Hosts}
\end{figure*}

Having compared theoretical and observational relations
of BHs in classical and pseudo-bulges, we here move on to discussing the
full demography of the predicted and observed BH populations via the
BH mass functions. While this type of comparison was pursued
several times in the Literature \citep[e.g.,][]{Lapi06,Marulli08},
it was never performed before taking care of differentiating classical and pseudo bulges.

To this purpose, we first recompute an empirical estimate of the local
BH mass function (BHMF).  A common way to obtain an estimate of the BHMF
is via convolution of the velocity dispersion function (VDF) of local
galaxies, with the \mbh$-\sigma$ relation (the convolution is between
the VDF and a Gaussian with dispersion equal to the intrinsic scatter in
the \mbh$-\sigma$ relation).  To this purpose, we use the latest
determination of the Hubble type-dependent VDF from \citet{Bernardi10},
coupled with a revised \mbh-\sis\ relation by Shankar et al. (2011, in
preparation) obtained from an updated sample of \citet{FerrareseFord},
i.e., $\log \mbhe = (4.57 \pm 0.47)\log \sigma+8.26\pm0.08$ with
intrinsic scatter $\eta=0.17 \pm 0.08$ (for early-type galaxies), and
$\log \mbhe = (4.99 \pm 0.38)\log \sigma+8.18\pm0.06$ with intrinsic
scatter $\eta=0.17 \pm 0.06$ for later-type galaxies (though similar
conclusions are derived adopting other estimates of the relation). The
black, dot-dashed line in Fig. 4 is the result of the convolution
including all Hubble types, with the dotted lines marking the 1-$\sigma$
error bars derived from Monte Carlo simulations that include statistical
uncertainties on the VDFs and scaling relations (including uncertainties
in the intrinsic scatter).  The estimated BH mass function is in good
agreement with previous estimates \citep[e.g.,][grey band]{SWM}.


However, \citet{Fisher11} recently suggested that pseudobulges may be
the dominant class of bulges at low masses. Moreover, as discussed
above, pseudobulges host significantly lower mass BHs than expected from
the \mbh-\sis\ relation.  The long-dashed, cyan line in Fig. 4 shows
the local BH mass function obtained by assuming that all bulges in Sb
galaxies are actually pseudobulges and thus follow the \mbh-\sis\
relation by \citet{Hu08}. (We do not claim that there is a separate
\mbh-\sis\ relation for pseudobulges; the \citet{Hu08} relation is used
here just for illustration.) The solid, cyan line is the extreme estimate
in which all bulges in Sb and Sa galaxies are pseudobulges. This exercise
shows that accounting for pseudobulges can easily decrease the BH mass density
by $\sim 15\%$, when correcting only Sb, up to $\sim 35\%$ in the
extreme case that all Sb/Sa galaxies are actually pseudobulges, while BH
number densities decrease by a factor of
$\sim 2-3$ below $\lesssim 10^8 \, \msune$ (these results do not depend much on the assumed scatter in the \citet{Hu08} relation).



The right panel of Fig. 4 compares the local BH mass function
corrected for pseudobulges (cyan area, obtained by combining the cyan
long-dashed and solid lines in the left panel), with the global BH mass
function at $z=0$ predicted by the B06 and G11 models (long-dashed,
black and solid, blues lines, respectively).  It is interesting to note that
\emph{both} models still predict too many classical (i.e., more massive
at fixed \mstar) BHs in the local Universe, thus tending to overpredict
the number of low mass BHs, in the range $10^7$ and $10^8$ \msun, with
respect to the empirical BH mass function corrected for pseudobulges
(below $10^7$ \msun, the number density continues to be overpredicted by
G11, but it is underpredicted by B06).

From this rather straightforward experiment
we thus conclude that if BHs in pseudobulges really are the most common type of BHs in the local Universe, then models still
need to properly account for such a large population (see also
\citealt[][]{Fontanot11,Shankar11}).  Recent estimates of the local BH
mass function \citep[e.g.,][]{Vika09} claim even lower number densities
of low mass BHs that would further exacerbate the tension with model
predictions. The shape of the local BH mass function
 at low masses is essential to properly constrain the total amount of sub-Eddington accretion
that characterized the cosmic evolution of BHs \citep{SWM2}.


\section{Discussion}
\label{sec|discu}

The comparison in Fig. 1 suggests that a tight link between bar
instabilities and BH growth at all times and in all conditions
may be disfavored.
More specifically, we have explored the outputs of the B06 model that couples
BH growth to mergers \emph{and} disk instabilities with both fuelling modes
tuned to match the median local scaling relation between
BHs and bulges masses in early-type galaxies.
We have then compared the model outputs against a
(recent) collection of data on BH masses in local
active and inactive galaxies of late-type morphological type,
mostly identified as pseudobulges,
that tend to be outliers with respect to the scaling relations
of BHs in earlier-type, classical bulges. The model struggles to
fully reproduce the latter selection of data.
In other words, the intrinsic scatter predicted by the B06 model
induced by differences in galaxy-to-galaxy evolutionary paths
is not sufficient by itself to match the recent
data on BHs in later type galaxies/pseudobulges.

The G11 model instead, triggering BHs only in mergers,
allows for significant, late bulge growth via bar instabilities with no
parallel fuelling in the central BH, thus naturally creating a larger
scatter in the BH-bulge mass relations.
Our results thus suggest that models in which late time disk instabilities, mainly characterized by the formation of \emph{bars},
are not closely related to BH growth, tend to generate a larger scatter in the \mbh-\mstar\ relation in better
agreement with the broad distribution of BHs we observe today.
This finding confirms prior expectations \citep[see, e.g.,][ and
references therein]{Kormendy11,Mathur11}.

We note here that several other galaxy/AGN evolutionary models based on BH fuelling
in bulges with no bar instabilities \citep[e.g.,][]{Granato04,Ciras05,Shankar06} have
also suggested additional possible sources
of the observed large scatter observed in BH scaling relations.
A break in BH scaling relations at low masses induced by inefficiency
of quasar compared to supernova feedback,
responsible for shaping the galaxy stellar mass function
\citep[e.g.,][and references therein]{Shankar06},
and different normalizations of the \mbh-\mstar\ relation
at different virialization epochs of the host haloes,
inevitably increase the predicted distribution of BH masses
at fixed bulge mass.

Our results, however, do not imply that BH fuelling triggered by disk instabilities is never at work.
Within the highly turbulent interstellar medium
of the protogalactic disks at high redshifts mentioned
in Sect.~\ref{sec|intro} \citep[e.g.,][]{Genzel11},
large clumps can form and migrate towards the centre via dynamical friction.
At variance with bars that can build up bulges gradually
on very long timescales, the latter type of instabilities are rapid and violent processes. Moreover, the clumpy disk instabilities may potentially form central spheroids structurally more similar to early-type bulges, as displayed by some recent numerical simulations \citep[e.g.,][but see also Inoue \& Satoh 2011]{Bournaud11}.
The clumpy inflows within high-$z$ disks have been recently
recognized to also play a non-negligible role in feeding the central BHs \citep{Frederic11}, possibly under the effect of continuous cold flow accretion \citep{DiMatteo11}.
In this framework, the turbulent disks at high redshifts seem to share similar effects to mergers, forming \emph{classical} bulges and proportionally growing central BHs \citep[e.g.,][and references therein]{Frederic11}.

Our results may equivalently reflect on a still not
fully appropriate treatment of BH fuelling
during bar-like disk instability events.
The way BH feeding follows
late bar instabilities may in fact be a non-trivial function
of other parameters such as gas fractions,
environment, etc... that may further
contribute to produce the local scatter we observe.

Our understanding of BH feeding is in fact still fuzzy,
especially at lower redshifts when gas reservoirs in galaxies
become poorer.
\citet{Peeples06} raised some doubts about a strong link
between bars and AGN activity.
They investigated the correlation between bar strength and circumnuclear dust morphology
for 75 galaxies finding that in many cases strongly barred galaxies showed
structures terminating in a circumnuclear starburst ring
at $\sim 10$ pc away from the galactic nucleus.
On the other hand, many Narrow Line Seyfert 1 galaxies appear as barred
systems and their AGNs are believed to be fuelled via bars \citep[e.g.,][]{Cren03,Orban11}.

\citet{CoelhoGadotti} recently claimed evidence for double
the fraction of AGNs in barred galaxies as compared to unbarred galaxies,
at least for low-mass bulges.
\citet{Oh11} have recently analyzed a large sample of barred and unbarred
galaxies extracted from the Sloan Digital Sky Survey concluding that
the correlation between bars and AGN activity (and SFR) is far from trivial
and also difficult to establish due to secondary correlations between
AGNs, galaxy morphology, etc...
They however point out that some connection between
AGN activity and bars seems to be present, at least in the local Universe.
In particular, they claim that galaxies having
longer bars appear to have more AGNs than their
counterparts at fixed stellar mass and morphology.

Indirect evidence for BH growth via bars may also come from higher redshifts.
\citet{Cisternas11} find the BH to total host stellar mass ratio to
be constant since redshift $z=0.9$. Given that their sample
is mainly composed by galaxies with disks, they claim that secular processes
might have fuelled both a central bulge and correlated BH in time thus preserving their
correlation down to $z=0$.
More generally, other additional sources of BH fuelling at later times are also
probably at work, such as winds from stars
\citep[e.g.,][and references therein]{Ciotti01,KauffmannHeckman,Cen11}.

Overall, evidence for some AGN activity in barred/pseudobulged systems may be
present. However, there is also mounting evidence that these
systems lie mostly below the relations characterizing classical bulges,
for both active and inactive galaxies. \citet{Beifiori12} have
recently revisited a number of BH scaling relations
in a revised sample of local galaxies with BHs, and claimed, in line with
\citet{Kormendy11}, that pseudobulges have little
or no correlation between BH mass and velocity dispersion. Thus,
it is not at all certain that BHs in non-classical bulges
should end up obeying similar scaling laws.
\citet{Orban11}, for example, suggest that Narrow Line Seyfert 1 may not be in any special phase
of their evolution, but simply are BHs growing slowly due
to their specific duty cycle and spin, mainly driven by secular processes in their pseudobulge hosts.

As discussed in Sect.~\ref{sec|BHMFpseudo}, \citet{Fisher11} recently suggested that pseudobulges may be
the dominant class of bulges at low masses. 
Thus, the number of outliers in the BH-galaxy classical relations
could be very high, impacting the demography of local BHs
by severely reducing the expected number density of low mass BHs in the local Universe
(similar thoughts were also put forward by \citealt{GadottiKauffmann} and \citealt{Greene10}).
This in turn could have a non-negligible repercussion
on accretion and merging models that try to tune
parameters such as the radiative efficiency and Eddington ratio distributions
by matching the full shape of the local BH mass function \citep[e.g.,][]{ShankarReview,Marulli08,SWM2}.

\begin{table}
  \caption{Galaxies hosting ``pseudobulges'' with known black hole masses}             
\begin{tabular}{c c c c}
  \hline
  Galaxy & Log \mbh\ [$M_{\odot}$] & Log \mstar\ [$M_{\odot}$] & Reference\\
  \hline
  NGC1068 &   6.92 &  10.60 & H09 \\
 NGC3079 &   6.40 &  10.24 & H09 \\
 NGC3393 &   7.49 &  10.62 & H09 \\
Circinus &   6.04 &   9.62 & H09 \\
 IC2560 &   6.46 &  10.31 & H09 \\
     MW &   6.61 &  10.11 & H09 \\
     MW &   6.63 &  10.28 & K11 \\
Circinus &   6.08 &  10.14 & K11 \\
 NGC1068 &   6.93 &  11.20 & K11 \\
 NGC1300 &   7.85 &  10.14 & K11 \\
 NGC2748 &   7.67 &   9.80 & K11 \\
 NGC2787 &   7.61 &   9.80 & K11 \\
 NGC3227 &   7.18 &   9.97 & K11 \\
 NGC3384 &   7.26 &  10.52 & K11 \\
 NGC4736 &   6.82 &  10.42 & K11 \\
 NGC4826 &   6.13 &  10.30 & K11 \\
 NGC7582 &   7.74 &  10.29 & K11 \\
 MS2254 &   6.60 &  10.25 & M11 \\
 RXJ1209 &   6.75 &  10.53 & M11 \\
 RXJ1117 &   7.33 &  10.49 & M11 \\
 RXJ2217 &   7.10 &  10.45 & M11 \\
 RXJ1702 &   7.34 &  10.53 & M11 \\
 KUG1136 &   6.89 &   9.91 & O11 \\
 MRK0042 &   6.22 &   9.91 & O11 \\
 Mrk0359 &   6.70 &  10.65 & O11 \\
 Mrk0382 &   6.59 &  10.26 & O11 \\
 Mrk0493 &   6.35 &  10.07 & O11 \\
 Mrk0766 &   7.14 &   9.61 & O11 \\
 Mrk0896 &   7.01 &  10.25 & O11 \\
 Mrk1044 &   7.20 &  10.03 & O11 \\
 NGC4748 &   6.95 &  10.07 & O11 \\
 Mrk0871 &   7.67 &  10.22 & O11 \\
 Mrk1126 &   6.90 &  9.80  & O11 \\
 Circinus&   6.23 &  10.26 & S11\\
 IC2560  &   6.64 &  10.94 & S11\\
 NGC1068 &   6.93 &  11.33 & S11\\
 NGC3079 &   6.40 &  11.02 & S11\\
 NGC3368 &   6.88 &  10.74 & S11\\
 NGC3489 &   6.78 &  10.07 & S11\\
 NGC3998 &   8.38 &  10.36 & S11\\
 NGC4258 &   7.58 &  10.98 & S11\\
 NGC4594 &   8.76 &  11.23 & S11\\
  \hline
  \end{tabular}
  \tablefoot{Details on the data listed in this Table can be found
  in Sect.~\ref{sec|data}. Here H09, K11, M11, O11, and S11 refer to \citet{Hu09}, \citet{Kormendy11}, \citet{Mathur11}, \citet{Orban11}, and \citet{Sani11}, respectively.}
  \label{table|models}
\end{table}

\begin{acknowledgements}
FS acknowledges support from a Marie Curie Fellowship. We thank the referee for a thoughtful report
that helped to significantly improve the presentation of the paper.
We thank N. Fanidakis, L. Ferrarese, F. Fontanot, P. Nair, R. Yates, and G. Zamorani
for interesting discussions, and E. Sani and G. Orban de Xivry for providing us with their data.
\end{acknowledgements}

\bibliographystyle{aa} 
\bibliography{aa.bbl}

\end{document}